%% file: part1
\documentstyle{europhys0} 
\input euromacr

\input epsf
\newcommand{\gs}{\sigma}

\newcommand {\gb} {\beta }

\begin{document}
\euro{}{}{}{}
\Date{}
\shorttitle{H.P. YING \etal QUNTUM XY MODEL}
\title{
Dynamic Monte Carlo Study of \\
the Two-Dimensional Quantum XY Model
}
\author{H.P. Ying\inst{1}, 
        H.J. Luo\inst{2}
        L. Sch{\"u}lke\inst{2}
        B. Zheng\inst{3}
        }
\institute{
     \inst{1} 
 Zhejiang Institute of Modern Physics, Zhejiang University,
 Hangzhou 310027,~ P.R. China\\
     \inst{2}
 Fachbereich Physik, Universit\"at-GH Siegen, D-57068 Siegen, 
 Germany\\
     \inst{3} 
 Fachbereich Physik, Universit{\"a}t Halle, D-06099 Halle, 
 Germany
 }
\rec{}{}
\pacs{
\Pacs{75}{10Jm}   {Quantized spin models}
\Pacs{75}{40Mg}   {Numerical simulation studies}
\Pacs{64}{60Ht}   {Dynamic critical phenomena}
      }

\maketitle
 
\begin{abstract} 
We present a dynamic Monte Carlo study of the
Kosterlitz-Thouless phase transition
 for the spin-$1/2$ quantum XY model in two-dimensions.
The short-time dynamic scaling behaviour is found and  
the dynamical exponent $\theta$, $z$ 
and the static exponent $\eta$ are determined at the 
transition temperature.
\end{abstract} 

%%%%%%%%%%%%%%%%%%%%%%%%%%%%%%%%%
%%%%%%%%% 

The existence and the nature of the phase transition in 
the quantum XY model is a long-standing problem.
In 1973, Kosterlitz and Thouless explained
what is now called  
the Kosterlitz-Thouless (KT) phase
transition in the classical XY model, in terms of topological 
order, characterized by an exponentially divergent 
spatial correlation length
and susceptibility  \cite{Kost73}.
General universality arguments suggest
that the same KT transition may occur in the quantum XY model
 \cite{Cull83, Loh85, Ding90}.
However, a quantitative determination of the critical exponents
and the transition temperature with Monte Carlo methods
is very difficult since one suffers from critical slowing down.

Due to the {\it exponential} divergence of the correlation length
at the transition temperature $T_{KT}$
and the fact that the system remains
critical below $T_{KT}$, 
numerical simulations of critical systems with
a KT transition are more difficult than those with a second order 
phase transition. The situation is even more severe
for quantum spin systems with a KT transition
 \cite{Ding90,Ding92}.
A standard approach to the quantum XY model
is the quantum Monte Carlo (QMC) method
where the Suzuki--Trotter transformation
is used to transform the quantum system to a classical one
 \cite{Suzu86,Raedt84}.
For the 2-dimensional spin-$1/2$ quantum XY model,
Loh, Scalapino and Grant first estimated 
the KT transition temperature between $T_{KT}=0.4-0.5$ 
 \cite{Loh85}.
The authors of refs.  \cite{Ding90, Ding92} 
improved the results with extensive QMC simulations on 
lattices up 
to
$128^2$. With a loop-cluster algorithm
 \cite{Ever93,Wiese94} which is often more efficient than
the conventional QMC methods, Harada and Kawashima
recently measured the helicity modulus 
for temperatures between 
$T=0.2-0.60$, and 
determined rather accurately
 the transition temperature $T_{KT} = 0.3423(2)$ 
on the lattice $64^2$  \cite{Kenji97}.
In this loop-cluster algorithm a loop is formed 
by spin-pairs on the
interacting plaquettes and all spins on the loop are flipped
simultaneously to overcome critical slowing down
  \cite{Wiese92}. However, up to now it is still
 difficult to determine the critical exponents in
equilibrium accurately. For example, the critical exponent $\eta$
has been given as $\eta=0.25 \pm 0.01$ in ref.  \cite{Ding90}, 
and $\eta=0.290 \pm 0.09$ or $\eta=0.276 \pm 0.014$ in ref.  
\cite{Ding92}.  
These values differ from each other.

On the other hand, in recent years much progress
has been made in critical dynamics. Traditionally it was believed 
that universal scaling behaviour
only exists in equilibrium or in the long-time regime of
the dynamic evolution. 
However, for {\it classical} magnetic systems,
it was discovered that universal scaling behaviour 
emerges already in the {\it macroscopic short-time regime}
\cite{Jans89,Huse89,li94,Schu95,gra95,li96,sta97,Okano972,Luo97}. 
Important is that new independent exponents
must be introduced to describe the dependence of the scaling
behaviour on the initial conditions,
or specify the scaling behaviour of special dynamic
observables. More interestingly, 
based on the short-time dynamic scaling form,
static exponents and the dynamic exponent $z$
originally defined in equilibrium or in the
long-time regime of the dynamic
evolution can be extracted already from the
universal short-time behaviour
 \cite {li96,zhe98,luo98}.
This provides a possible new way out of
critical slowing down.

In this paper,  we investigate whether there exists 
universal short-time dynamic scaling behaviour
in {\it quantum} spin systems, taking
the spin-1/2 quantum XY model in two dimensions
as an example. We determine
the new exponent $\theta$ and the dynamic exponent $z$
as well as the static exponent $\eta$ from
the power law behaviour of the observables
at the beginning of the time evolution.

The spin-1/2 quantum XY model is defined by the Hamiltonian
\begin{eqnarray}
\hat H =  -J\ \sum_{<ij>} [s_i^x \cdot s_j^x
+ s_i^z \cdot s_j^z ]~, 
\label{def_H} 
\end{eqnarray} 
where $<ij>$ stands for nearest-neighbour pairs 
on a 2-dimensional lattice, 
$s_i^x$ and $s_i^z$ are spin operators
defined on each lattice site, which can be expressed by
the Pauli matrices as
$(s^x,s^z)=\frac{1}{2}(\sigma^x,\sigma^z)$. 

By the checkerboard decomposition \cite{Suzu86,Mak91} 
with the Suzuki-Trotter formula we express the partition
function as
\begin{equation}
Z = \mbox{Tr}[\mbox{exp}(-\gb \hat H)] = \lim_{m \rightarrow \infty}
\mbox{Tr}[\mbox{exp}(-\epsilon {\hat H}_1)\mbox{exp}(-\epsilon {\hat 
H}_2)
\mbox{exp}(-\epsilon {\hat H}_3)\mbox{exp}(-\epsilon {\hat H}_4)]^m.
\label{trotter} 
\end{equation}
Here
$\gb =1/T$ is the inverse temperature, $\epsilon = \gb/m$
and the set of ${\hat H}_i$ arises from a decomposition of
 $\hat H$ as defined in ref.  \cite{Wiese94}.
Since the coupling constant $J$
can be absorbed into the temperature, we put $J=1$
in our later discussions.
Now we insert complete sets of eigenstates $|+1\rangle$ and 
$|-1\rangle$ 
of $\gs^z$ between each factor of $\mbox{exp}(-\epsilon 
{\hat H}_i)$ to map the 2-dimensional quantum system to an 
induced
(2+1)-dimensional
classical system with Ising-like variables $s(i,r)=\pm 1/2$,
\begin{equation}
Z  = \sum_{s(i,r)=\pm 1/2} 
\mbox{exp}[-S(\{s(i,r)\})]. ~~~~~~~~~~~~~~~~~~
\label{partition} 
\end{equation}
Here $r$ labels the slices in the artificial third dimension 
which has total $4m$ layers. 
$S(\{s(i,r)\})$ consists of the four-spin interaction
associated with the $r$-like plaquettes,
which are called interacting plaquettes.
Each interacting plaquette
is bounded by four non-interacting plaquettes to form a
checkerboard lattice  \cite{Wiese92}. The interacting
plaquette weights are determined 
by the products of the transfer matrix 
\begin{eqnarray}
 \mbox{exp}[-S_p(s_1, s_2; s_3, s_4)]
=<s_1,s_2 \vert \mbox{exp}[\frac {\epsilon}{4}
( \gs_{i}^x \cdot \gs_j^x 
+\gs_{i}^z \cdot \gs_j^z)]\vert s_{3},s_{4}>,
\label{weight}
\end{eqnarray}
for each plaquette configuration $C_p(s_1,s_2;s_3,s_4)$ with 
$s_1$ and $s_2$ locating on a $r$th-slice, 
and $s_3$ and $s_4$ on the $(r+1)$th-slice  \cite{Wiese92}. 
Actually $S(\{s(i,r)\})=\sum_p S_p(s_1, s_2; s_3, s_4) $ and
the sum is over all possible interacting plaquetts $p$.
A calculation of the transfer matrix 
$exp[-S_p(s_1, s_2; s_3, s_4)]$
leads to the form
\begin{eqnarray}
\mbox{exp}{(\frac {\epsilon}{4})} 
\pmatrix
{ \cosh (\frac{\epsilon}{4})  & 0 & 0 & \sinh (\frac {\epsilon}{4})\cr
0 & \mbox{exp}(-\frac {\epsilon}{2}) \cosh (\frac {\epsilon}{4}) 
  & \mbox{exp}(-\frac {\epsilon}{2}) \sinh (\frac {\epsilon}{4}) & 0\cr
0 & \mbox{exp}(-\frac {\epsilon}{2}) \sinh (\frac {\epsilon}{4})
  & \mbox{exp}(-\frac {\epsilon}{2}) \cosh (\frac {\epsilon}{4}) & 0 \cr
\sinh (\frac {\epsilon}{4}) & 0 & 0 & \cosh (\frac{\epsilon}{4}) }~.~~~~
\label{matrix_xz}
\end{eqnarray}
The states of the spin-pair ($s_1, s_2$) for the row index and 
($s_3, s_4$) for the column index
are arranged in a sequence of $(+,+)$, $(+,-)$, $(-,+)$, $(-,-)$.
The elements of the matrix (up to an overall constant)
can be interpreted
as the Boltzmann weights of the configurations 
for the induced classical spin system.
The zero elements indicate that only 
a part of the spin configurations is allowed.

Updating schemes based on the transfer matrix
must satisfy the ergodicity and detailed balance condition.
Due to the facts that any allowed configurations can be achieved by 
flipping pairs of spins on plaquette edges from an allowed
configuration and 
each spin is shared by two interacting plaquettes \cite{Mak91}, 
it follows
immediately that one should flip a {\it closed loop} of spins,
{\it i.e.} change the signs of all spins on the loop simultaneously.
Practically, we adopt the updating procedure introduced in refs. 
 \cite{Mak91,Kawa96} in our simulations.
A complete Monte Carlo sweep updates:
1) all {\it spatial } non-interacting 
plaquettes bounded by four interacting plaquettes; 
2) all $r$-like loops formed by three connected
non-interacting plaquettes
bounded by eight interacting plaquettes;
3) all Polyjakov lines in the $r$-direction;
4) all global loops generated by connecting diagonal 
spin-pairs in the interacting plaquettes and vertical spin-pairs in
the non-interacting plaquettes for all $(i,r)$-like planes.

We first consider a dynamic relaxation process
starting from an ordered state ($m_0=1$)
 \cite{Luo97,zhe98}. In this 
dynamic process, it is believed that at the transition
temperature $T_{KT}$ or below,
there exists a dynamic scaling form, {\it e.g.} for the $k$th moment of 
the 
magnetization, 
\begin {equation}
M^{(k)}(t)=b^{-k \eta /2} M^{(k)} (b^{-z} t, b^{-1}L),
\label {3_2}
\end {equation}
which sets in right after a {\it microscopic} time scale
$t_{mic}$. This time scale can depend on initial conditions, 
algorithms
or other microscopic details.
In simulations of classical spin systems
$t_{mic} \sim 100$ Monte Carlo time steps
 is observed.
If a Monte Carlo time step is considered to be a typical
microscopic time unit, this is reasonable.

Starting from all spins up, 
we have performed the simulations with the Metropolis
algorithm at the transition temperature
$T_{KT}=0.3423$  \cite {Kenji97}. Lattice sizes are taken to
be $L^2 \times 4m  = 64^2 \times 120$ and 
$L^2 \times 4m = 32^2 \times 120$.
Samples for average are over $1\ 000$ for $L=64$ and
$2\ 000$ for $L=32$.
Statistic errors are estimated by
dividing the samples into three groups.
We measure the magnetization defined as
 \begin{eqnarray}
M(t) =\frac 1{L^2\times 4m}<\sum_{(i,r)}s(i,r,t)>
\label{magt}
\end{eqnarray}
and its second moment. From the scaling form in (\ref {3_2}),
it is easy to deduce that the magnetization decays
by a power law
\begin {equation}
M(t) \sim t ^{-\eta /2z}.
\end {equation}
Such a power law decay has actually been known for a long time
in the long-time regime of the dynamic evolution,
but now it is expected to hold also in the macroscopic short-time
regime.
To determine the dynamic exponent $z$ independently
and further confirm the short-time dynamic scaling,
we introduce a dynamic Binder cumulant 
$U(t,L)=M^{(2)}/M^2-1$. Simple finite size scaling analysis shows 
\begin {equation}
U(t,L) \sim t^{d/z}.
\end {equation}
This behaviour of the Binder cumulant is a typical behaviour
in the short-time regime,
where the non-equilibrium spatial correlation length
is very small.

\begin{picture}(6,6)(20,100)
\epsfysize=6cm
\setlength{\unitlength}{.6cm}
\put(0,-5){{\epsffile{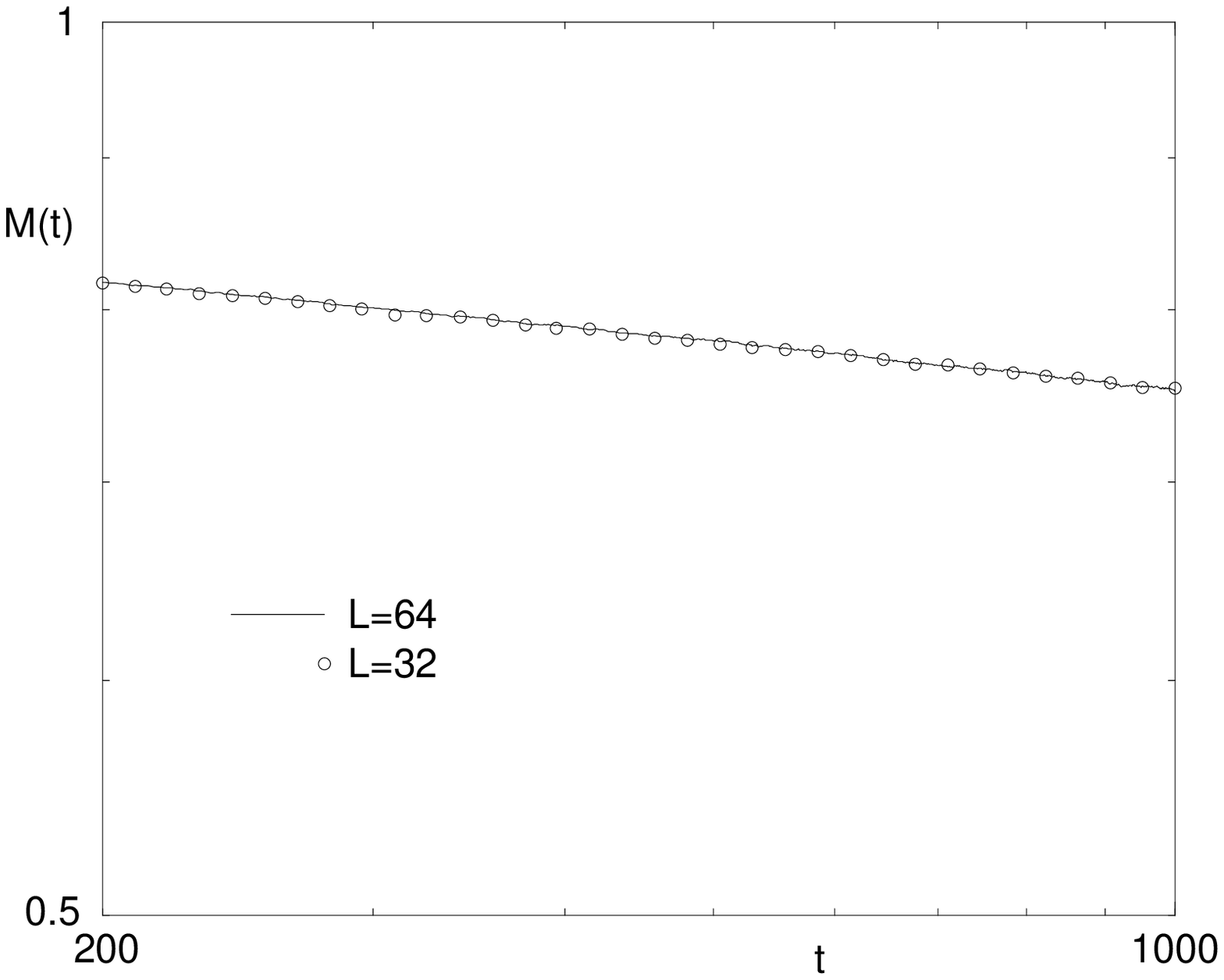}}}
\epsfysize=6cm
\setlength{\unitlength}{.6cm}
\put(12,-5){{\epsffile{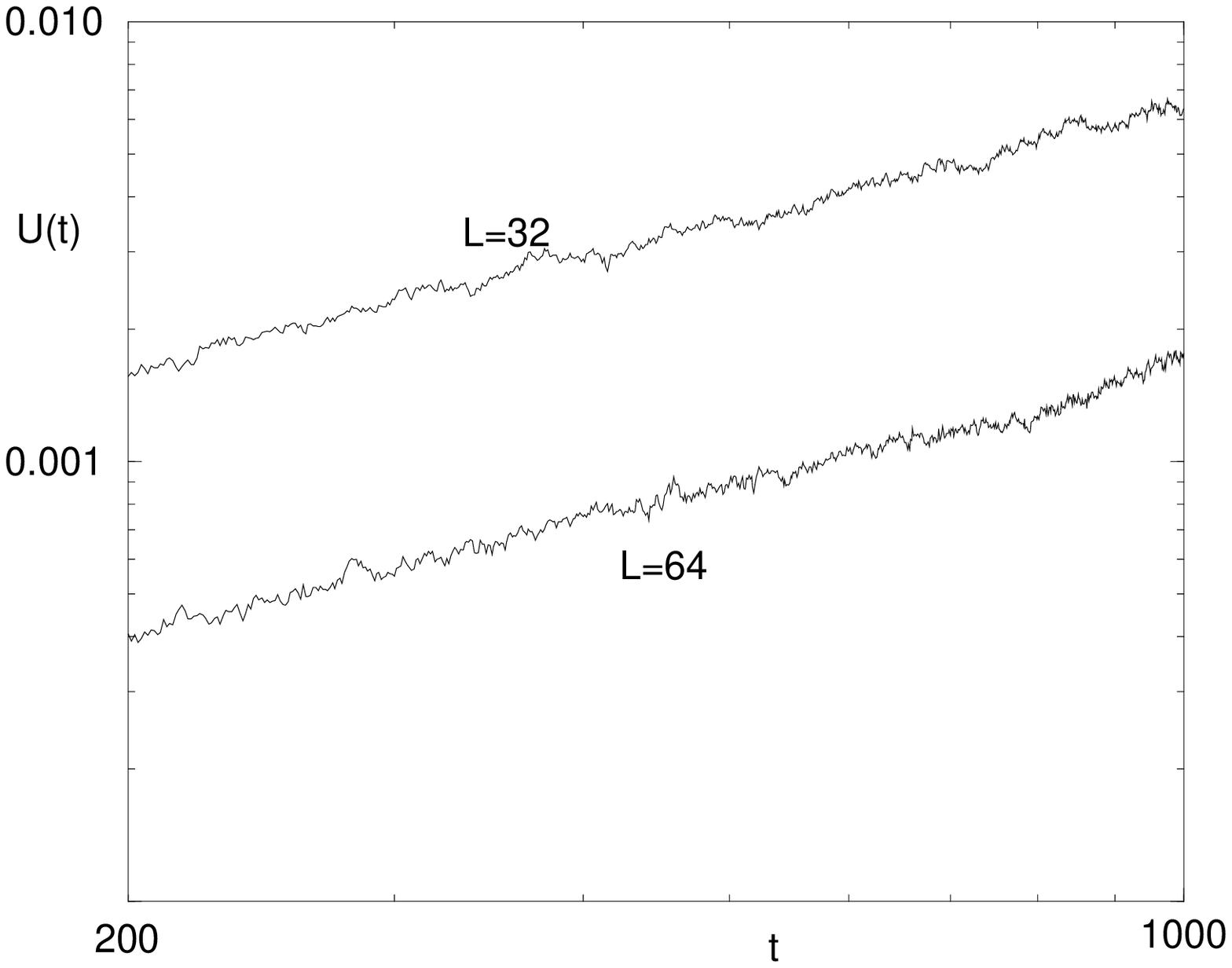}}}
\put(5,-5){Fig.~\ref{M_gs}}
\put(17,-5){Fig.~\ref{C_gs}}
\end{picture}

\vspace{5.5cm}
\begin{figure}[h]
\caption{ The time evolution of the magnetization
for $L=64$ and $L=32$ by starting from ordered state
is plotted in double-log scale.
}
\label{M_gs}
\end{figure}
\vspace{-1.5cm}
\begin{figure}[h]
\caption{ The time evolution of the cumulant 
for $L=64$ and $L=32$ by starting from ordered state
is plotted in double-log scale.
}
\label{C_gs}
\end{figure}

In fig. \ref{M_gs} the time evolution of $M(t)$ is plotted in
double-log scale. The solid line is that for $L=64$ and the 
circles
for $L=32$. One can see that until 1000 Monte Carlo
steps, there is no visible 
finite size effect for these lattice sizes.
Here the microscopic time scale 
$t_{mic}\sim 100$. 
In the figure we have skipped data for $t$ smaller than $200$.
From the slope 
of $M(t)$ in the time interval [200, 1000] 
we measure the critical exponent $\eta /2z = 0.0522(6)$ for $L=64$
and for $L=32$ we get $\eta /2z =0.0518(7)$. Within the statistic 
errors, 
they coincide.

From the measured 
$M(t)$ and $M^{(2)}(t)$ we calculate the Binder cumulant 
$U(t)$ which is plotted in fig. \ref{C_gs}.  
It is interesting that 
the curves of $U(t)$ exhibit short-wave fluctuations,
which have not been observed 
in classical spin systems  \cite {Luo97,zhe98}.
However, the short-wave fluctuations do not
affect so much the long-wave behaviour and a 
clear power law behaviour is seen in the figure.
From the slopes of the curves,
we measure the critical exponent 
$d/z = 0.85(3)$ for $L=64$ and $d/z =0.86(3)$ for $L=32$.
With the results of $L=64$, 
we get the critical exponent $z=2.35(8)$.
Taking $z$ as input, from the measured  value of $\eta 
/2z=0.0522(6)$
we obtain
$\eta =0.245(8)$.  
Compared with the results from simulations
in equilibrium, our value of the exponent $\eta$
supports $\eta=0.25\pm 0.01$ in ref.~ \cite{Ding90}
but is smaller than 
$\eta=0.29\pm 0.09$ or $\eta=0.276\pm 0.014$ given in
ref.~ \cite{Ding92}.
This suggests that the exponent $\eta$ in this quantum XY model
takes a classical value $\eta=1/4$.

In the short-time regime of the dynamic evolution,
there are plenty of new phenomena.
To describe all these phenomena, the static exponents
together with the dynamic exponent $z$ 
are in general not sufficient.
Another interesting and important
dynamic process is the relaxation starting from
a high temperature initial state  
with a small initial magnetization $m_0$.
It has been shown analytically and observed numerically
in classical spin systems, that 
after a microscopic time scale $t_{mic}$, 
the magnetization undergoes surprisingly 
a power law initial increase  \cite{Jans89,li94,Schu95}
\begin {equation}
M(t) \sim m_0 t^\theta.
\label {3_1}
\end {equation}
The exponent $\theta$ is a new independent
critical exponent, which can not be expressed in terms of other
known exponents. 

In order to measure the critical exponent $\theta$, 
we must prepare an initial state of random configurations
but with a small initial magnetization $m_0$.
In the high temperature limit of the transfer matrix
in eq. (\ref{matrix_xz}), the four elements proportional to
$\sinh (\frac {\epsilon}{4})$ vanish
and only the four equal-weighted non-zero diagonal elements 
remain,
which correspond to the interacting plaquettes of $C_p(s_1,s_2;
s_3,s_4) = (+,+;+,+)$, $(-,-;-,-)$, $(+,-;+,-)$ and $(-,+;-,+)$.
Therefore, we first put randomly $s_i^z = \frac 12$ with
 a probability of $p=(m_0 + 1)/2$
 and $s_i^z = - \frac 12$ with a probability of $1-p$ respectively
on the first layer $(r=1)$ of the $L^2 \times 4m$ lattice.
Then we copy this configuration to all other $r-1$ layers.
This procedure gives the required initial random configuration.

\begin{figure}[h]\centering
\epsfysize=6cm
\epsfclipoff
\fboxsep=0pt
\setlength{\unitlength}{.6cm}
\begin{picture}(6,6)(1,2.5)
\put(-2,1){{\epsffile{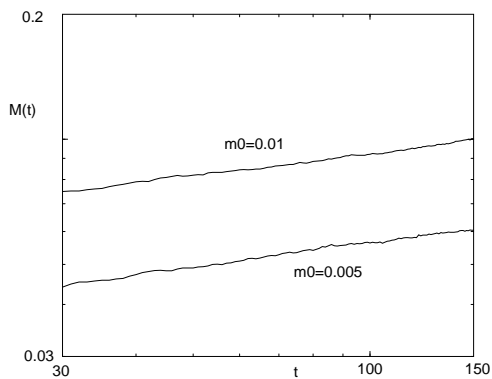}}}
\end{picture}
\caption{ The time evolution of the magnetization
for $L=64$ with $m_0=0.01$ and $m_0=0.005$
is plotted in double-log scale.
}
\label{M_rs}
\end{figure}

The simulations have been performed with a lattice size $L=64$ 
and total samples over $4\ 000$. It is well known that the 
microscopic time scale $t_{mic}$ in this case is small,
typically less than $20$ Monte Carlo time steps
 \cite {li94,Schu95,Okano972,zhe98}.
Therefore we stop updating at $t=150$.
In fig. \ref{M_rs},
the time evolutions of $M(t)$ are plotted in log-log scale.
After about 30 Monte Carlo steps, 
the curves show a nice power law behaviour.
From the slope of the curves, we measure the critical exponent
$\theta =0.18(1)$ for $m_0=0.01$ and $0.18(2)$ for $m_0=0.005$.
Rigorously speaking, the exponent $\theta$ is defined in the limit
$m_0 \rightarrow 0$. However, within statistical errors
our results for $m_0=0.01$ and $0.005$ show already no 
difference.
Therefore an extrapolation of $\theta$ to
the limit $m_0 = 0$ is not necessary here.
\vspace{1cm}
In conclusions, 
for the first time we have investigated the short-time critical
dynamics of the quantum spin system taking the two-dimensional
spin $1/2$ quantun XY model as an example.
Universal power law behaviour of the magnetization and
the Binder cumulant is found and the new critical
exponent $\theta$, the dynamic exponent $z$ and the
static exponent $\eta$ are determined at the transition
temperature $T_{KT} =0.3423$.
The measured exponent $\eta=0.245(8)$ agrees very well with
$\eta=0.244(5)$ obtained for the classical XY model  \cite{Luo97}
and also with the theoretical prediction $\eta=0.25$.
However, the exponent $\theta=0.18(2)$ and the dynamic exponent
$z=2.35(8)$ are apparently different from
$\theta=0.250(1)$ and $z=1.96(4)$ for the classical XY model
 \cite{Okano972,Luo97}.
These results indicate that the quantum XY model and the 
classical XY model are in a same {\it static}
universality class but in a different {\it dynamic}
universality class. As an alternative approach, our simulations
pave a way for a systematic application
of the short-time dynamic scaling to quantum
spin systems. 

For quantum spin systems, besides the stochastic relaxational
dynamics there exists a true dynamics described by the
Heisenberg equation of motion. It is challenging
whether the short-time dynamic scaling can be found 
also there.

\vspace{0.15in}
{\it Acknowledgement}: 
Work supported in part by the Deutsche 
Forschungsgemeinschaft; DFG Schu 95/9-1 and SFB 418. 
The author HPY is also grateful to the support 
by the NNSF of China under Grant No. 19575039. 

%%%%%%%%%%%%%%%%%%%%%%%%%%%%%%%%%
%%%%%%%%%%%%%%%%%%%%%%%%%%%%%%%%%%%
%%%%%%%%%

\baselineskip=12.0pt
%\newpage 

\end{document}

%% file: euromacr.tex
\def\etal{{\hbox{{\tenit\ et al.\/}\tenrm :\ }}}

\newif\ifboo \boofalse

%% file: part1.bbl
\begin{thebibliography}{99}
\bibitem{Kost73}
 J.M. Kosterlitz and D.J. Thouless, J. Phys. C{\bf 6}(1973)1181;\\
 J.M. Kosterlitz, {\it ibid}, C{\bf 7}(1974)1046.
\bibitem{Cull83} 
 J.J. Cullen and D.P. Landau, Phys. Rev. {\bf B27}(1983)297. 
\bibitem{Loh85} 
 E. Loh, D.J. Scalapino and P.M. Grant,  Phys. Rev. B {\bf 
31}(1985)4712.
\bibitem{Ding90} 
H.-Q. Ding and M.S. Makvi\'c, Phys. Rev. B{\bf 42}(1990)6827.
\bibitem{Ding92} 
 H.-Q. Ding, Phys. Rev. B{\bf 45}(1992)230, and references therein. 
\bibitem{Suzu86}
 M. Suzuki, Prog. Theor. Phys. {\bf 56}(1976)1454; J. Stat. Phys. 
 {\bf 43}(1986)833; \\R.M. Fye, Phys. Rev. B{\bf 33}(1986)6721.
\bibitem{Raedt84} 
 H.De Raedt, and A. Lagendijk, Phys. Lett. {\bf 104A}(1984)430; 
Z. Phys. B{\bf 57}(1984)209.
\bibitem{Mak91} 
M.S. Makvi\'c and H.-Q. Ding, Phys. Rev. B{\bf 43}(1991)3562. 
\bibitem{Ever93} 
 H.G. Evertz, M. Marcu and G. Lana, Phys. Rev. Lett. {\bf 
70}(1993)875.
\bibitem{Wiese94} 
 U.-J. Wiese and H.-P. Ying, Z. Phys. B{\bf 93}(1994)147.
\bibitem{Kenji97} 
 Kenji Harada and Naoki Kawashima, Preprint of 
cond-mat/9702081.
\bibitem{Wiese92} 
 U.-J. Wiese and H.-P. Ying, Phys. Lett. A{\bf 168}(1992)143;\\
 H.-P. Ying and F. Chen, Phys. Lett. A{\bf 208}(1995)356.
\bibitem{Kawa96}  
 N. Kawashima, J. Stat. Phys. {\bf 82}(1996)131.
\bibitem{Jans89} 
 H.K. Janssen, B. Schaub and B. Schmittmann, Z. Phys. {\bf 
B73}(1989)539. 
\bibitem{Huse89} 
 D.A. Huse, Phys. Rev. {\bf B40}(1989)304. 
 \bibitem{li94} 
 Z.B. Li, U. Ritschel and B. Zheng, J. Phys. {\bf A27}(1994)L837.
\bibitem{Schu95} 
 L. Sch\"ulke and B. Zheng, Phys. Lett. {\bf A204}(1995)295. 
\bibitem{gra95} 
P. Grassberger, Physica {\bf A 214} (1995)547.
 \bibitem{li96} 
 Z.B. Li, L. Sch\"ulke and B. Zheng, Phys. Rev. {\bf E53}(1996)2940.
\bibitem{sta97} 
 D. Stauffer, Physica {\bf A 244} (1997) 344.
\bibitem{Okano972} 
 K. Okano, L. Sch\"ulke, K. Yamagiishi and B. Zheng, 
 J. Phys. {\bf A30}(1997)4527. 
\bibitem{Luo97} 
 H.J. Luo, and B. Zheng,  Mod. Phys. Lett. {\bf B11}(1997)615. 
\bibitem{zhe98} 
 B. Zheng, {\it Monte Carlo Simulations of Short-time Critical 
Dynamics},
Halle Uni., 1998, review article, to be published in Int. J. Mod. Phys. 
B.
\bibitem{luo98} 
H.J. Luo, L. Sch\"ulke and B. Zheng, Phys. Rev. Lett. {\bf 81}(1998)180. 
\end{thebibliography}
